# Mespotine-RLE-basic v0.9

# An overhead-reduced and improved Run-Length-Encoding Method

20.01.2015

Meo Mespotine
trss.mespotine.de – mespotine@mespotine.de

Abstract

Run Length Encoding(RLE) is one of the oldest algorithms for data-compression available, a method used for compression of large data into smaller and therefore more compact data. It compresses by looking at the data for repetitions of the same character in a row and storing the amount(called run) and the respective character(called run_value) as target-data.

Unfortunately it only compresses within strict and special cases. Outside of these cases, it increases the data-size, even doubles the size in worst cases compared to the original, unprocessed data.

In this paper, we will discuss modifications to RLE, with which we will only store the run for characters, that are actually compressible, getting rid of a lot of useless data like the runs of the characters, that are uncompressible in the first place.

This will be achieved by storing the character first and the run second. Additionally we create a bit-list of 256 positions(one for every possible ASCII-character), in which we will store, if a specific (ASCII-)character is compressible(1) or not(0).

Using this list, we can now say, if a character is compressible (store [the character]+[it's run]) or if it is not compressible (store [the character] only and the next character is NOT a run, but the following character instead).

Using this list, we can also successfully decode the data(if the character is compressible, the next character is a run, if not compressible, the next character is a normal character).

With that, we store runs only for characters, that are compressible in the first place. In fact, in the worst case scenario, the encoded data will create always just an overhead of the size of the bit-list itself. With an alphabet of 256 different characters(i.e. ASCII) it would be only a maximum of 32 bytes, no matter how big the original data was.

Many image/audio/video-formats who apply Standard-RLE(FLAC, TIFF, etc), could benefit from Mespotine-RLE heavily by getting rid of the negative side-effects of Standard-RLE. Even data-compression programs that use RLE as main compression-method or as a pre-processor, could be improved by Mespotine-RLE.

## 1. Introduction

RLE is a comprehension-method and it basically works, like a shopping list. If you want to buy 4 bananas, you probably do not write “banana, banana, banana, banana”. You comprehend the list by writing “4 bananas” instead. By that, you need less space for your shopping list: you compressed the information in a way, that the original information(“banana, banana, banana, banana”) is easy to recover from the compressed(“4 bananas”) information.

Run Length Encoding works quite the same way. That means, it compresses by comprehending characters in the original data, that are stored repeatedly in a row in the original data.. To do this, we count the appearances of a certain character in the data. After that we encode it by storing how often this character shall be repeated(known as *run*) and the character itself(known as *run-value*)[2][1].

For example: AAA becomes 3A, where the 3 is the *run*(indicating this specific character was stored 3 times in the original data), and the A is the *run-value*(indicating, the specific character we deal with right now is the A).

If we comprehend the following original-data: BBBBAAOPPOOOOP = 14 characters
the encoded data looks like this. 4B 2A 1O 2P 4O 1P = 12 characters.

We saved 2 characters compared to the original-data → data is compressed by 2 characters

When decoding, we read from the encoded data the *run* and then the *run-value.* After that, we store the *run-value* for *run*-times until we decoded and by that restored the original data.

4B → BBBB
2A → AA
1O → O
2P → PP
4O → OOOO
1P → P

The decoded(decompressed) data is: BBBBAAOPPOOOOP

The downsides of this method are, that two characters in a row (like the AA or the PP in the example above) never create compression, as the encoded data is of the same size as the original-data. Even worse, single characters (like the first O and the last P in the example above), that needed only one byte in the original data, also get an additional *run* during the encoding-process; although this *run* does just indicate, that this specific character appears only once.

In the latter case, the encoded data becomes twice the size of the original, unencoded data (O → 1O, P → 1P). In worst-case-scenarios, this could create encoded data, that is twice the size of the original-data. One might be tempted to think “Let's just write the *run* only for characters, that are repeating at least three times, not for those appearing only twice or once!”

Unfortunately, if we do that, we loose predictability with RLE, as in computers, characters are stored with numbers(i.e. with ASCIII, an A is stored with a 65, B with 66, etc) and in the encoded data, the *runs* are also stored with numbers

If we throw away some *runs*, we run into problems like in the following, encoded data, as seen by the internals of the computer: 65 66 65 66 65 66

Is it: 65 B 65 B 65 B ?
Is it: 65 B A B 65 B ?
Is it: A B 65 B A B ?
Is it: A B A B A B ?
Is it: … ?

It is not clear, as we can not certainly say, which is the *run* and which is the *run-value*, both could be possibly appearing here. Therefore we MUST keep the order and store a *run* AND a *run-value* for every character appearing, even if it is for a character appearing only once. Otherwise, we might get confused with uncertainty and too many possibilities, as the next character could be interpreted as *run* **or** as *run-value*, **or** even **both**.And such confusion is only acceptable within lossy compression methods.

Standard-RLE is lossless.

So, does this mean, we need to accept this as a given? Isn't there a chance of getting rid of the *runs* for characters not compressible at all in the first place? And can't we get rid of the worst-case-scenario of encoded data, twice the size of the original data?

The answer to all these three questions is: There's a way of dealing with these problems. And we are going to discuss this in the next chapters in detail.

## 2. Mespotine RLE (Basic)

Before we start with the method itself, there are some basic differences between Standard RLE and Mespotine RLE-basic that we need to discuss first.

### 2.1 Idea

The biggest downside within classic RLE is rising from a tiny, but crucial problem: We tend to save a lot of data that we do not need for actual compression[2]. Therefore, we store useless data, despite the fact that it is, well: useless.

Where can we find the useless data? Well, certainly not in the *run-value*, as this is the information we definitely need for recreating the original data. So we need to have a look at the *run*, which we even store for *run-values*, that actually do not produce compression at all.

So the first change with Mespotine-RLE-basic is, we put the more important *run-value* first and the secondary important *run* second.

Uncompressed: AAAABBBBCCDDE
Standard – RLE: 4A 4B 2C 2D 1E
Mespotine – RLE: A4 B4 C2 D2 E1

Now we reversed the order, so what do we gain from it? Well: predictability. As we always need the *run-value*, it is the most important data in the encoding process. So we store it first. Now, all we need is a simple logic that decides for us, if the next character in the data is to be interpreted as a *run* or the next *run-value.* With that, we only need to store *runs*, that benefit us one way or another.

So the question arising from it: How is this logic actually working? And what do we need to make it work?

### 2.2 The Comp_Bit_List

To differentiate between characters that produce compression and those who don't, we need some kind of a reminder. In our case, it is the Comp_Bit_List, which is a bitlist with 256 entries(one for each ASCII-Character). Every entry could be set to 0 (uncompressible character) or set to 1(compressible character). So every character that is marked as compressible in our list will be encoded with RLE, the rest stays the way it is.

But how do we know which character is compressible and which is not?

We simply count all appearances of a specific character in the source-data and compare them with their encoded counterparts.

First we go through the data for the character with ASCII-code 0 and check, if encoding it using RLE would compress this specific character or not. This is done easily by just counting the compression-efficiency with the following rules:

a) If the character(in the current *run,* that we have analyzed*)* appears 3(+x) times in a row, we add 1(+x) to the variable “counter”

b) If the character(in the current *run,* that we have analyzed) appears 2 times in a row, we add 0 to the variable “counter”

c) If the character(in the current *run,* that we have analyzed) appears only 1 time in a row, we subtract 1 from the variable “counter”

Go on counting all the character-appearances, until you have reached the end of the original-data.

After analyzing all appearances of this specific character in the original-data, we take a look at the variable "counter":

1) If the variable "counter" is positive, we can successfully compress this character. The number of bytes we can save by applying RLE to this character is the number we have stored in "counter".
2) If the variable "counter" is 0, then this character will stay the same amount of characters, no matter if we apply RLE or not, no compression achieved.
3) If the variable "counter" is negative, then we have no compression for this character at all. Even worse: applying RLE makes the data for this character even bigger. To calculate, how bigger, just make the value stored in "counter" positive, and you know the number of characters that would be added to the encoded data, when you apply RLE to this specific character.

If the specific character is compressible, store in the accompanying entry of the Comp_Bit_List for this ASCII-character a 1, if it is not compressible, you should store a 0. (That means, if you checked the character A and it is compressible, the entry for ASCII-Character 65 within the Comp_Bit_List is set to 1)

After that, repeat the procedure with the next ASCII-characters(first 1, then 2, then 3, …, then 253, then 254, then 255).

Lets have a look at an example. Imagine, we have an alphabet of 4 characters in the data only: A, B, C, D. The original-data is as follows: AAAABBCCCDB

Next we create a Comp_Bit_List with 4 entries for this data. The first entry is for the A, the second for the B, the third for the C and the fourth for the D.

Now let's have a look at which character is compressible, using the rules above.

A comes 4 times in a row:Rule a: "counter" would be 1(+1) → 2 Bytes (positive → compression).
B comes 2 times in a row:Rule b: "counter" would be 0 → 0 Byte
B comes 1 time in a row :Rule a: "counter" would be -1 (0-1) → -1 Byte (negative → no compression)
C comes 3 times in a row:Rule a: "counter" would be 1(+0) → 1 Byte (positive → compression)
D comes 1 time: Rule c: "counter" would be -1 → -1 Byte (negative → no compression)

Now we set all the entries in the Comp_Bit_List. We set 1 for the characters that are compressible, and 0 for all the characters that are not compressible. The Comp_Bit_List would be as follows: 1010

In detail: the A($1^{st}$ entry) and C($3^{rd}$ entry) are compressible: each 1. B($2^{nd}$ entry) and D($4^{th}$ entry) are not compressible: each 0.

Let's do another example: AAABBBAACDAAAABDB

Analyzing A: The first batch of A is 3 characters (Rule a:*1* character saved), the second batch of A is 2(Rule b: *0* character saved), the third batch of A is 4 (Rule a: *2* characters saved). Now lets see, if the A is compressible: *1*+*0*+*2*=3 characters saved. The number(3) is positive, therefore the A is compressible.
→ $1^{st}$ Comp_Bit_List-entry must be set to 1

Analyzing B: The first batch of B is 3 characters(Rule a:*1* character saved), the second batch of B is 1(rule c: -*1* character saved), the third batch of B is 1(rule c: -*1* character saved). Is B compressible? *1*+(-*1*)+(-*1*)=-1. The number(-1) is negative, therefore the B is NOT compressible. → $2^{nd}$ Comp_Bit_List-entry set to 0

Analyzing C: The first batch of C is 1 character (Rule c: -*1* character saved). No other batch of C in the data. Now lets see if C is compressible: -*1*=-1. The number(-1) is negative, therefore the C is not compressible. → $3^{rd}$ Comp_Bit_List-entry set to 0

Analyzing D: The first batch of D is 1 character (Rule c: -*1* character saved). The second batch of D is 1 character (Rule c: -*1* character saved). Now let's see if D is compressible: *-1-1*=-2. The number(-2) is negative, therefore the D is not compressible. → 4th Comp_Bit_List-entry is 0

Now, lets create the Comp_Bit_List (1 for compressible, 0 for uncompressible characters): The A(1st entry) is compressible, therefore 1. All other three characters are not compressible, therefore 0.

The final Comp_Bit_List is 1000

With such a list, which contains an entry for every ASCII-character that could appear (max 256 in total), we can store which character is compressible and which is not. After that, we can check, if a specific character could be compressed or not. And, as we only need one bit for every entry to store such data, the whole list is only 256 bits in length (32 bytes).

### 2.3 Encoding

The idea is simple: We read the original-data, character by character, as usual with Standard-RLE. But, every time we read a new character, we take a look into the Comp_Bit_List, if the specific character is compressible at all or not. If the accompanying entry is set to 1, we apply RLE by storing *run-value* and the *run*. If the character is not compressible(the accompanying entry is set to 0), we just store the character as *run-value*, without(!) a *run.*

Lets take the two examples from the chapter before:

| | | |
|---|---|---|
| Data | : AAAABBCCCDB | = 11 characters= 88 bits |
| Comp_Bit_List | : 1010 | |
| Standard RLE | : 4A 2B 3C 1D 1B | = 10 chars= 80 bits |
| Mespotine-RLE | : A4 BB C3 D B | = 8 chars+Comp_Bit_List(4 bits)= 68 Bits! |

We read the first A in the original-data and checked with the Comp_Bit_List, if the A is compressible or not. The 1st Comp_Bit_List-entry is set to 1, therefore the A is compressible, so we can apply RLE to it: AAAA → A4

We read the first B in the original-data and checked, with the Comp_Bit_List, if the B is compressible or not. The 2nd Comp_Bit_List-entry is set to 0, therefore it is not compressible, so we store it the way it is: B → B

We read the second B in the original-data and checked, with the Comp_Bit_List, if the B is compressible or not. The 2nd Comp_Bit_List-entry is set to 0, therefore it is not compressible, so we store it the way it is: B → B

We read the first C in the original-data and checked with the Comp_Bit_List, if the C is compressible or not. The 3rd Comp_Bit_List-entry is set to 1, therefore the C is compressible, so we can apply RLE to it: CCC → C3

We read the D in the original-data and checked, with the Comp_Bit_List, if the D is compressible or not. The 4th Comp_Bit_List-entry is set to 0, therefore it is not compressible, so we store it the way it is: D → D

We read the third B in the original-data and checked, with the Comp_Bit_List, if the B is compressible or not. The 2nd Comp_Bit_List-entry is set to 0, therefore it is not compressible, so we store it the way it is: B → B

As you could see: With Mespotine-RLE applied, we only stored *runs* for the characters A and C. The B and D however, were stored without a *run*, therefore we saved the space of 2 characters, compared to Standard-RLE. Adding the size of the Comp_Bit_List added 4 bits, therefore, we saved 12 bits altogether with Mespotine-RLE, compared to Standard-RLE

Now let's take a look at the other example:

| | | |
|---|---|---|
| Data | : AAABBBAACDAAAABDB | = 17 characters = 136 bits |
| Comp_Bit_List | : 1000 | |
| Standard RLE | : 3A 3B 2A 1C 1D 4A 1B 1D 1B | = 18 chars = 144 bits |
| Mespotine-RLE | : A3 BBB A2 C D A4 B D B | = 14 chars+Comp_Bit_List(4bits)=116 bits! |

Of course, the more data you want to encode, the higher compression-ratio you may achieve. But if you can't achieve compression with any of the characters in the original data, the worst thing that could happen is, that you add the size of the Comp_Bit_List at the beginning of the "encoded" data(256 bits of bits set to 0) for all ASCII-characters(you would never store *runs* in such a case). Which is much less, than the worst-case-overhead with Standard-RLE.

So if you can compress the original data by at least 33 bytes(the size of the comp_bit_list+1 byte of "actual compression"), your data becomes smaller, as we do not need to store more useless *runs* than absolutely necessary.

**2.4 Decoding**

Decoding is more or less the same procedure as the encoding, only reversed. We read the Comp_Bit_List, in which we can see, if a character is compressible or not.

After that we read the data, character by character (or better *run-value* by *run-value).*

Check if the first *run-value* is compressible (take a look in our Comp_Bit_List. If the accompanying entry is set to 1, it is compressible. If set to 0, it is not compressible). If the *run-value* is compressible, the next character must be interpreted as *run,* if the *run-value* is not compressible, the next character must be interpreted as the next *run-value.*

Repeat it, until you are finished.

The first example above is processed like this:
The Comp_Bit_List is: 1010 (1st entry A, 2nd entry B, 3rd entry C, 4th entry D)
The compressed data is: A4 BB C3 D B

Read the first *run-value*(A). The A is compressible(1st entry Comp_Bit_List is set to 1).
Therefore the next character must be interpreted as *run*(4 times). → AAAA
Read the next *run-value*(B). The B is not compressible(2nd entry Comp_Bit_List is set to 0).
Therefore the next character must be interpreted as *run-value* → B
Read the next *run-value*(B). The B is not compressible(2nd entry Comp_Bit_List set to 0).
Therefore the next character must be interpreted as *run-value* → B
Read the next *run-value*(C). The C is compressible(2ndt Comp_Bit_List entry set to 1).
Therefore the next character must be interpreted as *run*(3). → CCC
Read the next *run-value*(D). The D is not compressible(4th entry Comp_Bit_List set to 0).
Therefore the next character must be interpreted as *run-value* → D
Read the next *run-value*(B). The B is not compressible(2nd entry Comp_Bit_List is set to 0).
Therefore the next character must be interpreted as *run-value* → B

The decoded data is: AAAABBCCCDB
We successfully decoded and by that restored the original-data.

**2.5 Encoding *runs* longer than 256 efficiently - the *long_run***

Sometimes, we stumble over the situation of a *long_run*: we want to encode *runs,* that are longer, than the value-range of the *run* allows. In our case, that means, a *run* of more than 256 characters.

In Standard-RLE, we handle this situation quite simple: We start another encoded *run* by writing the next *run* and after that the next *run_value*(which is actually the same *run_value* as the previous one). This is inefficient, as we already know, that it is the same *run_value* we want to encode here and waste the space of a byte, for storing information we already know.

In Mespotine-RLE, we do things differently with *long_runs*.

To differentiate between a normal *run* and a *long_run*, we use escape-values in the *run*: We use the 255 and 256.

A *run* of 255 means: The *run_value* must be stored 255 times, the next value is the next *run_value*.

A *run* of 256 means: The *run_value* must be stored 255 times BUT: the next value is a *run*(!), that we add to the preceding *run*. If the next *run* is again 256, it is again a *long_run*. But if the *run* has a value smaller than 256, then it is the last *run* for the *run_value* of this *long_run*. That means, the next value we read is the next *run_value*.
Note: We use a value-range from 1-256 in this chapter!

For example: [A] [256] [5] = store A for 260(255+5) times (← *long_run*).
[B] [256] [256] [256] [255] = store B 1020 times (← *long_run*).
[A] [256] [1] = store A 256 times (← *long_run*).
[A] [255] [B] [256] [20] [C] [20] = store A 255 times,
B 275 times (← *long_run*),
C 21 times.

Note the difference: [256] → *long_run* (255+more to come)
[255] → normal *run* of 255

By that, we only store *run_values* for *runs*, where we need to know a corresponding *run_value.* But for *long_runs*, it is the same *run_value* anyway, no need to store useless *run_values* which would result in loosing compression efficiency.

But there is one downside with this method: *runs* of 256, can't be stored the way we did before : [A] [256] , but rather [A][256][1]. I personally think, that the gain for *long_runs* is better than the loss of efficiency for "rare" cases of "real"-*runs* of 256. So in the end we will benefit a lot from this approach.

This changes the way, we need to calculate the Comp_Bit_List a little by adding one rule to the three we already have; a "sub-rule" to rule a):

Rule a.1) If the current *run* we have analyzed is a multiple of 256, we subtract 1/256 (the size of the *long_run*-Escapevalue)from the variable "counter" for every multiple of 256 we have encountered (256=1/256, 512=2/256, ..., 65025=255/256, 65280=256/256, etc), to get exact numbers in compression achieved by this specific character.

Note: the multiples that we calculate with are 256, 512, 768, etc

In the encoding process: if we encounter a *long_run* (more than 255 characters), we store the *run_value* and after that a *run* of 256, which indicates a "real"-*run* of 255+more to come *runs*). After that, we only store *runs* until we have a *run,* that is only 255 or smaller (value-range 1-255).

The decoding is similar: when a *run* is 256, the next character must be interpreted as a next *run* of the current *run_value*, If the *run* is smaller(1-255), the next character is to be interpreted as the next *run_value*.

For example:

[D] [1] [C] [256 ← *indicator of a long_run*] [25] [A] [255][T][20]

**2.6 Bit Level Application**

You can also successfully apply Mespotine-RLE on bit-level-basis(for monochromatic images or faxes and such). With Standard-RLE[2] you encode it with the seven least significant bits storing the *run*(0-127), the most significant bit storing the *run_value*(0 or 1).

In Mespotine-RLE you modify it as I described it in chapter 2.1: you switch around the order. the least significant bit is the *run-value*(0 or 1), the seven most significant bits are the *run*. The Comp_Bit_List is only two bits long(one for the 0, one for the 1).

Calculating the Comp_Bit_List is a bit different on bit-level. You count the number of bits of the *run_value* 0 AND the number of *runs* the 0 has in the data. You do the same thing with *run_value* 1.

If the *number_of_bits*(0)>(*number_of_runs*(0)*8), then the *run_value* 0 is compressible, if not, it is not compressible.

The same with the *run_value* 1:
If the *number_of_bits*(1)>(*number_of_runs*(1)*8), then the *run_value 1* is compressible, if not, it is not compressible.

Now, you apply Mespotine-RLE as usual: you read one bit, you have a look into the Comp_Bit_List if it's compressible(1) or not(0). If compressible, the next seven bits are the *run*, if it is not compressible, the next bit is a *run-value.*

With that, you can decide, if one color(i.e. black) of a monochromatic image is compressible or not and do not need to store potential useless *runs* for that color.

The idea of storing a long_run without additional *run_values* could also be applied. That would mean, that a *run* of 127 is 127 times the *run_value*, a *run* of 128 is 127 times the *run_value* + more additional *runs* to follow.

The structure for a *run* of *160* of **zeros** would be: …] [**0**][*128*][*33*] […

Again, we loose a little compression efficiency because of the escape-value 128 for *runs*, which takes away the value 128 for “real”-*runs* of 128. To reflect that, we need to change the way we calculate the Comp_Bit_List the following way:

If *number_of_bits*(*run_value*)>(*number_of_runs*(*run_value*)*8)+*long_run**1/128)
→ *run_value* is compressible, else *run_value* is uncompressible.

*long_run* means here: the number of times you would use the *run* of 128, the escape value.

## 3. Flowcharts and Structures

In this chapter, I programmed a flowchart version of Mespotine-RLE. Unlike the previous chapters, where I used the value-range from 1-256 for a run, I'm using the value-range from 0-255, as would be necessary in real-computer implementations. The escape-values for a *long_run* therefore are 255 (a *run* of 254+more to follow) and 254(for a normal *run* of 254).

### 3.1 Creating Comp_Bit_List

Terms : Source → Original Data-Source
Variables : *counter* //counter for the length of a *run*
*ASCIIValue* //The ASCII-value of a character read from source-data
*ASCIIValue_new* //ASCII-value of next character read from source-data
*ASCIIValue_*counter[255] //final *run*-counter list for every character [0-255]
*i*=0 // a simple loop-count-variable
*Comp_Bit_List*[255] //The Comp_Bit_List for every character[0-255]

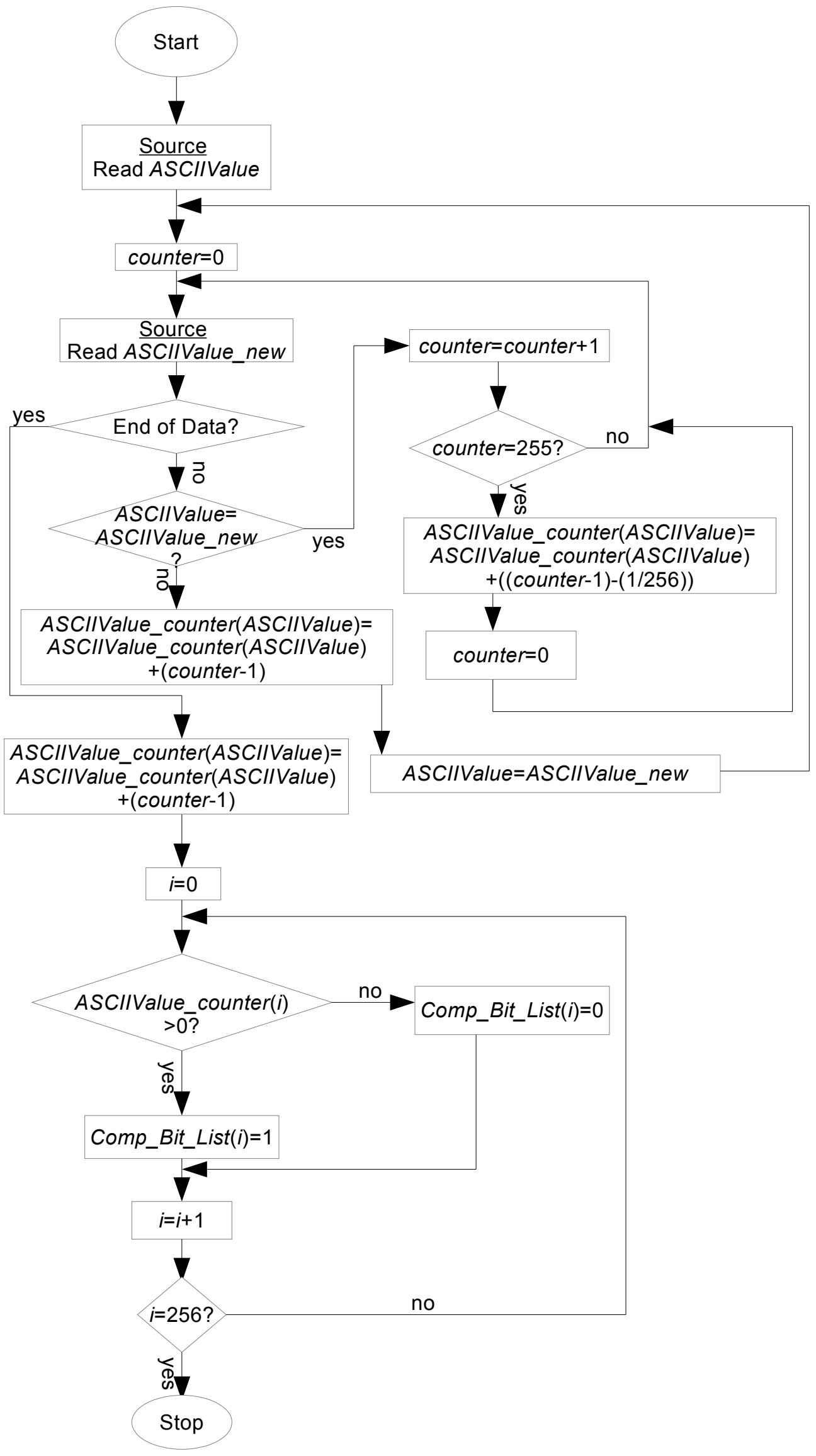

**3.2 Encoding**

Terms : Source → Original Data-Source
Target → The target for the encoded data

Variables : *run_counter* //counter for the length of a *run*(0-255)
*ASCIICode* //The ASCII-value of a character read from source-data
*ASCIICode_new* //ASCII-value of next character read from source-data
*Comp_Bit_List*[255] //The Comp_Bit_List for every character[0-255]
*long_run* //if we currently process a *long_run*(1) or not(0)

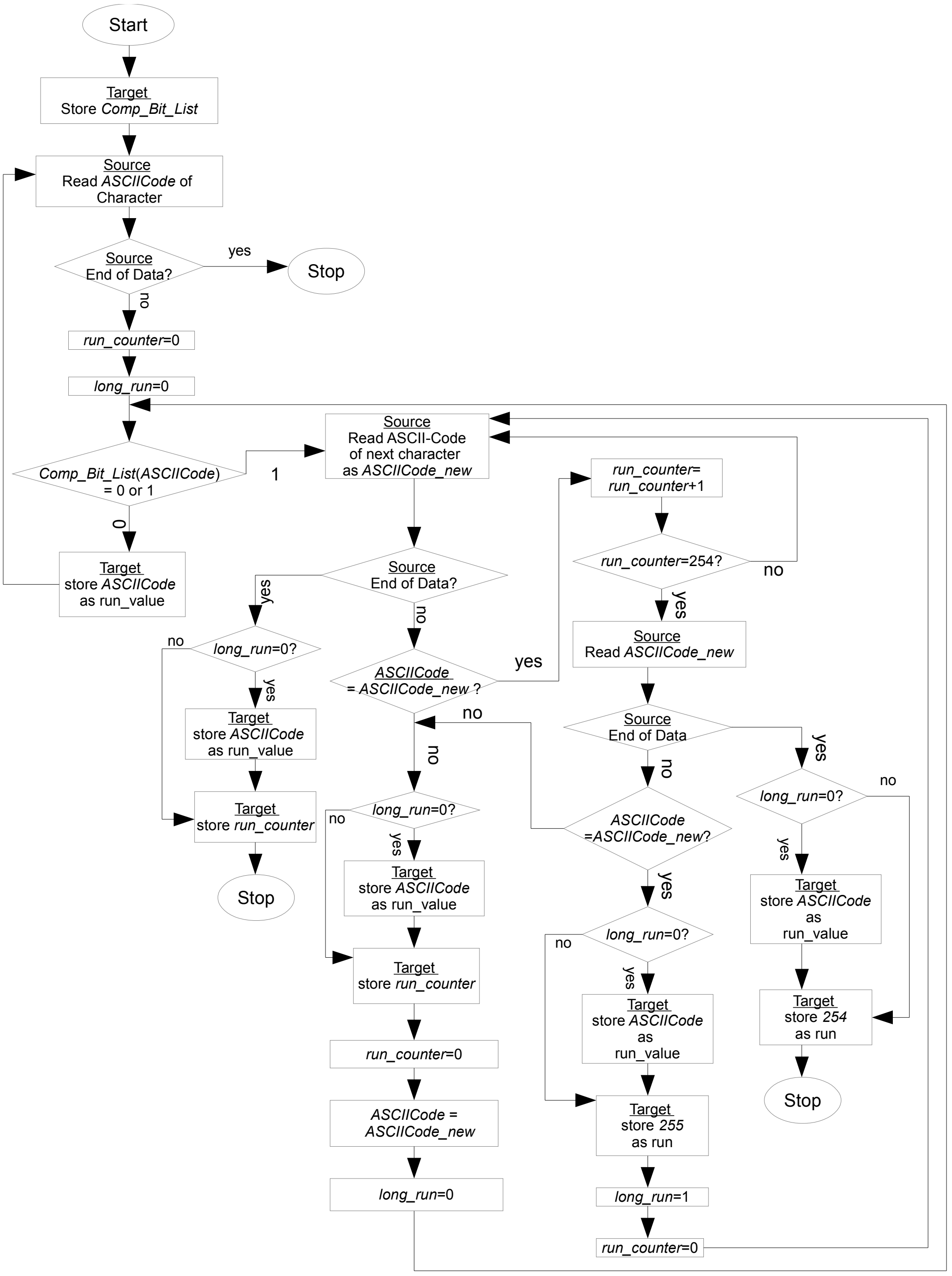

**3.3 Decoding**

Terms : <u>Source</u> → Original Data-Source
<u>Target</u> → The target for the encoded data

Variables : *run_value* //The ASCII-value of a character read from source-data(value-range 0-255)
*run* //the *run* of a compressible character read from source-data(0-255)
*Comp_Bit_List*[255] //The Comp_Bit_List for every character[0-255]

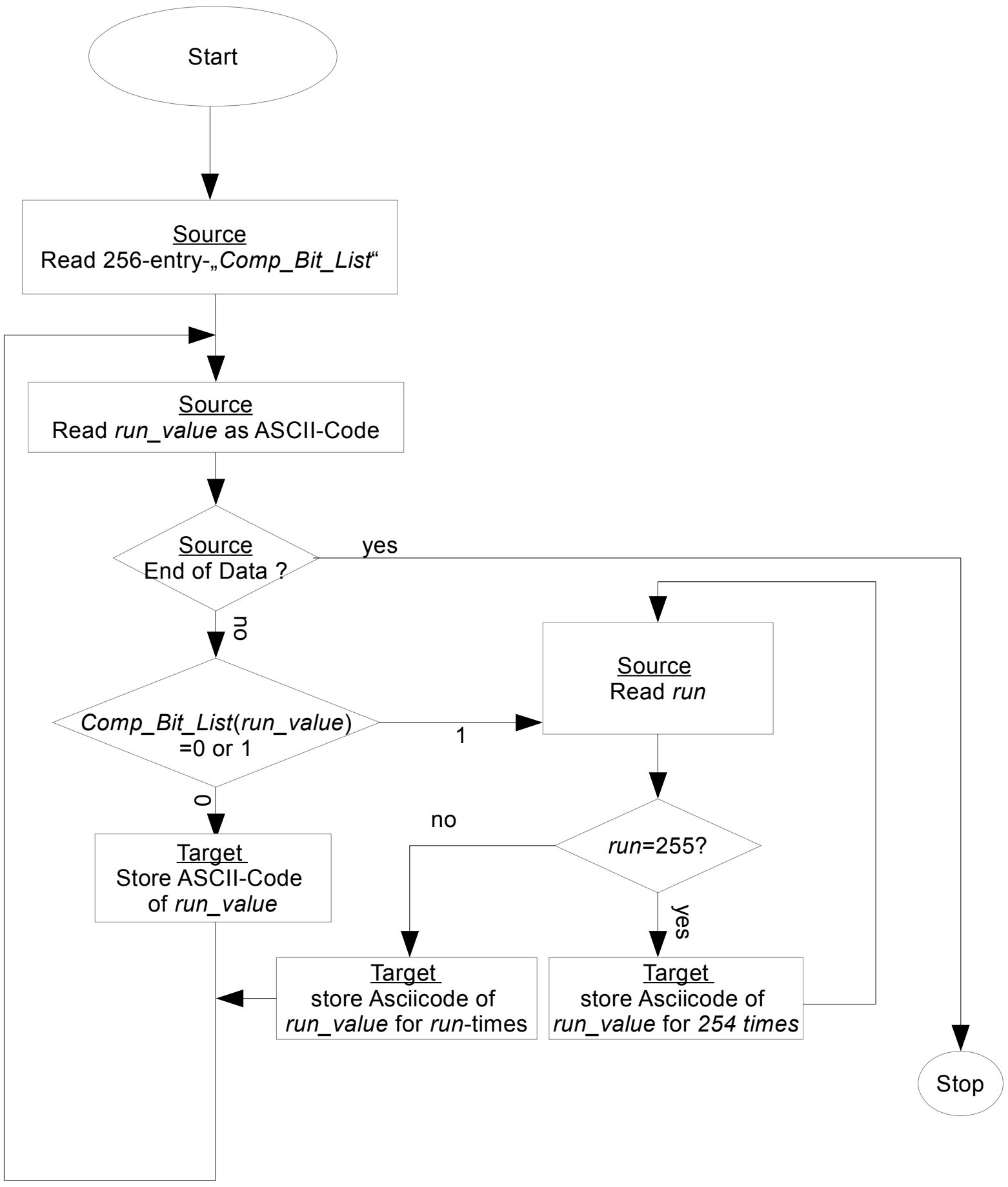

### 3.4 The File Format and logic overview

The encoded file follows the following structure:

[Comp_Bit_List ]→ [Encoded Data]

The Comp_Bit_List follows the following structure:

[entry for character 00] [entry for char 01] [entry for char 02] [entry for char 03] ...
... [entry for char 254] [entry for char 255]

The encoded data follows the following structure:

a) a uncompressible *run_value* is encoded
[*run_value*] [*run_value*]...
b) an compressible *run_value* is encoded
[*run-value*] [*run<255*][*run_value*]
c) an encoded run of maximum 255
[*run-value*] [*run*=255] [*run-value*]
d) an encoded run of longer than 255 (here a run of 260)
[*run-value*] [*run*=256(1-255)] [*run*=5(256-260)] [*run-value*]

## 4. Conclusion

As we could see, the Comp_Bit_List-concept and the new decision-logic applied to RLE produced much better compression-results in examples and test-cases, many of them weren't compressible before with Standard-RLE.

For *runs* longer than 255, we save 8 bits for each instance of an encoded *run* of that kind by just including an escape-value within a *run* that tells us, if we have a *long_run*, or not. As we already know, which *run_value* we have for the current *run*, we don't need to store it again and again, as with Standard-RLE. By that, we got rid of a lot useless *run-values*.

In worst-case-situations, the encoding does not produce doubled-sized-encoded-data anymore, but rather 32 bytes overhead only.

Because of that, algorithms, fileformats, video/audio-codes, that already apply Standard-RLE, could benefit a lot from using Mespotine-RLE, gaining more efficiency by getting rid of useless overhead created by Standard-RLE without significant loss in speed during encoding/decoding.

Additionally, unlike other methods, like PackBits[1] or Escape-Code-attempts like Tsukiyama's[3] method or similar, it is easier to implement, yet more efficient than these others in most cases.

The downsides of Mespotine-RLE are, that single-character-*runs* within compressible characters still create a lot of useless overhead, that could be eliminated. This is better achieved in Tsukiyama's method. Maybe a combination of Mespotine-RLE and Tsukiyama's method or even the Packbits-attempt is a possibility (i.e. the current and the next 3 of the "compressible" A's are unencodeable(**-3**): A10 B A **-3** B A B A B A10 compared to Mespotine-RLE-basic: A10 B A1 B A1 B A1 B A10).

Therefore, there is still a lot room for improvements on RLE. Some of them will be the subjects covered in my next papers.

## 5. License

## 6. Acknowledgments

Big thanks go to Mr Wagner from the LDS Schulungszentrum - Potsdam/Germany, as well as to Ulrich Grude and Volkmar Miszalok from Beuth Hochschule Berlin/Germany for teaching me basics in compression theory as well as the necessary tools to go ahead into the wild realms of data-compression.

I'd also like to thank Gal Schkolnik and Annelie Wendeberg for reading and commenting on earlier drafts of this paper.

Last, but not least, I'd love to thank the writers and contributors of the newsgroup comp.compression, who helped me a lot with understanding in depth what compression of data is, and what it is not.

I owe you a lot....

## 8. Author's Profile

Méô Mespotine has studied informatics at LDS-Brandenburg in Teltow/Germany 2000-2003, as well as at Beuth Hochschule für Technik in Berlin/Germany from 2004-2008. He is currently researching in compression-theory on a freelance-basis.

Despite other areas of research, his IT-interests focus mostly on Compression Algorithms, new analog and digital Human-Machine-Interface-concepts and Webtechnologies of Refinding WebContent fast and easy.

trss.mespotine.de
mespotine@mespotine.de

## A.Appendix
### A.1 Comparison Mespotine-RLE with Standard-RLE

1) **Original-Data : AAAAAAAAABBBBBBCDCDCDCDCDCDCDCD** = 31 characters=248 bits
Comp_Bit_List : 1100
Standard RLE : 9A 6B 1C 1D 1C 1D 1C 1D 1C 1D 1C 1D 1C 1D 1C 1D 1C 1D
= 36 characters=288 bits (ratio: 116.13%)
Mespotine-RLE : A9 B6 CDCDCDCDCDCDCDCD
= 20 characters+Comp_Bit_List=164 bits (ratio: 66.13%)

2) **Original-Data : ABCAABBCCDAAABBBCCCDAAAABBBBCCCCD** = 33 characters=264 bits
Comp_Bit_List : 1110
Standard RLE : 1A 1B 1C 2A 2B 2C 1D 3A 3B 3C 1D 4A 4B 4C 1D
= 30 characters =240 bits (ratio: 90.9%)
Mespotine-RLE : A1 B1 C1 A2 B2 C2 D A3 B3 C3 D A4 B4 C4 D
= 27 characters+Comp_Bit_List=220 bits (ratio: 83.3%)

3) **Original-Data : AAABBBCCCDDDEEE** = 15 characters=120 bits
Comp_Bit_List : 11111
Standard RLE : 3A3B3C3D3E = 10 characters =80 bits (ratio: 66.67%)
Mespotine-RLE : A3B3C3D3E3 = 10 characters+Comp_Bit_List=85 bits (ratio: 70.83%)

4) **Original-Data : AABBACCCDAABBB** = 14 characters=112 bits
Comp_Bit_List : 0110
Standard RLE : 2A 2B 1A 3C 1D 2A 3B = 14 characters =112 bits (ratio: 100%)
Mespotine-RLE : AA B2 A C3 D AA B3 = 12 characters+Comp_Bit_List=100 bits (ratio: 89.286%)

5) **Original-Data : AAABCDAAACBDAAADBC** = 18 characters=144 bits
Comp_Bit_List : 1000
Standard RLE : 3A 1B 1C 1D 3A 1C 1B 1D 3A 1D 1B 1C = 24 characters =192 bits (ratio:133.3%)
Mespotine-RLE : A3BCDA3CBDA3DBC = 15 characters+Comp_Bit_List=124 bits(ratio: 86.11%)

6) **Original-Data : ABCDABCDABCDABCD** = 16 characters=128 bits
Comp_Bit_List : 0000
Standard RLE : 1A 1B 1C 1D 1A 1B 1C 1D 1A 1B 1C 1D 1A 1B 1C 1D
= 32 characters=256 bits (ratio: 200%)
Mespotine-RLE : ABCDABCDABCDABCD
= 16 characters+Comp_Bit_List=132 bits (ratio: 103.125%)

As you can see in these comparisons, in most cases, where Standard-RLE produced no or negative compression, the Mespotine-RLE algorithm creates compression. Only within the sixth example, we have data that is bigger than the original-data, but by the size of the Comp_Bit_List only(in that case, only 4 bits bigger!), while example 3 creates slightly more negative compression compared to Standard-RLE, but also just bigger by the size of the Comp_Bit_List.

**An improvement ranging from 11%(example 4) up to 97%(example 6) in efficiency could be achieved in most of these examples with the different approach of Mespotine-RLE, compared to the compression-ratios of Standard-RLE.**

## A.2 Comparison with some familiar methods of RLE improvements

I applied to all the examples from A.1 some known and common methods, with whom RLE has been improved in the past.

Tokuhiro Tsukiyama with others[3] improved it by including an "Escape-character"-attempt: at least two occurring characters indicate a *run* of at least 2: AAAA becomes AA2 (AA is the indicator, the *run* of 2 tells, how often this character needs to be repeated).

This has some benefits, but also other downsides: compression is only achieved by 4 repeating characters in a *run*. Three create same sized data, two are making it bigger. On the other hand, characters who appear only once, are stored the way they are.

AAABCDAABCDAAAAA becomes AA1BCDAA0BCDAA3

Another method, used by the Packbits-Algorithm[1], is working the following way: the value-range for the *run* is split into three value-areas:

-127 to -1: how often is the character repeated

0 to 127: how many of the next characters shall not be encoded

-128: do nothing

With that, only *runs* up to 127 are possible. On the other hand, you could encode "*runs*" of uncompressible characters with just adding one "*run*"-byte, unlike Standard RLE, where every character would get a *run*-Byte.

AAABCDAABCDAAAAA becomes -3A7BCDAABCD-5A

In the following overview, I'm going to apply all four methods (Standard-RLE, Tsukiyama, Packbits and Mespotine-RLE) to the examples from chapter A.1. Ratios are in comparison to the size of the original-data:

| | | |
|---|---|---|
| **AAAAAAAAABBBBBBCDCDCDCDCDCDCDCD** | | **=31 characters** |
| StandardRLE: | 9A6B1C1D1C1D1C1D1C1D1C1D1C1D1C1D | =33chars (ratio: 116.13%) |
| Tsukiyama: | AA7BB4CDCDCDCDCDCDCDCD | =22chars (70.97%) |
| Packbits: | -9A-6B15CDCDCDCDCDCDCDCD | =21chars (67.74%) |
| MespotineRLE: | A9B6CDCDCDCDCDCDCDCD | =20chars+4Bits(66.13%) - *<u>smallest</u>* |
| **ABCAABBCCDAAABBBCCCDAAAABBBBCCCCD** | | **=33 characters** |
| StandardRLE: | 1A1B1C2A2B2C1D3A3B3C1D4A4B4C1D | =30chars(90.91%) |
| Tsukiyama: | ABCAA0BB0CC0DAA1BB1CC1DAA2BB2CC2D | =33chars (100%) |
| Packbits: | 2ABC-2A-2B-2C0D-3A-3B-3C0D-4A-4B-4C0D | =28chars(84.4%) |
| MespotineRLE: | A1B1C1A2B2C2DA3B3C3DA4B4C4D | =27chars+4Bits(83.3%)- *<u>smallest</u>* |
| **AAABBBCCCDDDEEE** | | **=15 characters** |
| StandardRLE: | 3A3B3C3D3E | =10chars (66.67%) - *<u>smallest</u>* |
| Tsukiyama: | AA1BB1CC1DD1EE1 | =15chars (100%) |
| Packbits: | -3A-3B-3C-3D-3E | =10chars (66.67%) - *<u>smallest</u>* |
| MespotineRLE: | A3B3C3D3E3 | =10chars+5Bits (70.83%) |
| **AABBACCCDAABBB** | | **=14 characters** |
| StandardRLE: | 2A2B1A3C1D2A3B | =14chars (100%) |
| Tsukiyama: | AA0BB0ACC1DAA0BB1 | =17chars (121.43%) |
| Packbits: | -2A-2B0A-3C0D-2A-3B | =14chars (100%) |
| MespotineRLE: | AAB2AC3DAAB3 | =12chars+4Bits(89.29%) - *<u>smallest</u>* |
| **AAABCDAAACBDAAADBC** | | **=18 characters** |
| StandardRLE: | 3A1B1C1D3A1C1B1D3A1D1B1C | =24chars (133%) |
| Tsukiyama: | AA1BCDAA1CBDAA1DBC | =18chars (100%) |
| Packbits: | -3A2BCD-3A2CBD-3A2DBC | =18chars (100%) |
| MespotineRLE: | A3BCDA3CBDA3DBC | =15chars+4Bits (86.11%) - *<u>smallest</u>* |
| **ABCDABCDABCDABCD** | | **=16 characters** |
| StandardRLE: | 1A1B1C1D1A1B1C1D1A1B1C1D1A1B1C1D | =32chars (200%) |
| Tsukiyama: | ABCDABCDABCDABCD | =16chars(100%) - *<u>smallest</u>* |
| Packbits: | 15ABCDABCDABCDABCD | =17chars (106,25%) |
| MespotineRLE: | ABCDABCDABCDABCD | =16chars+4Bits(103,125%) |

As seen in these examples, Tsukyama has some compression-benefits, when encoding *runs* of 4+x characters, as it creates, in most cases, smaller or at least the same data-size than the original-data was. We also have the benefit, that *runs* of single characters do not need to be encoded at all, but are stored the way they are.

But the improvements are at the cost of compressing *runs* of 3 characters, which is impossible now(the encoded *run* is still 3 bytes). Additionally, when *runs* of pairs occur, the data becomes bigger (like in the 4th example), eating up the improvements in this method.

Packbits however is even better, making compression like Standard-RLE encoding situations possible (*runs* of 3 characters=smaller, *runs* of 2 characters=same size).

Unfortunately, we can only encode *runs* with maximum number of 127 occurrences. We also need to include at least one *run*-byte for signaling unencodeable *runs*, which itself makes data bigger(like in the 6th example). And this signaling also has the limitation of a maximum *run* of 128 characters.

That means, after a 128 single character-"*run*" or a normal *run*, we need to include another *run*-byte if the (un-)encodeable *run* still continues. This is an improvement over Standard-RLE, but still eats up a lot of the possible compression.

Mespotine-RLE is improving on both of these areas, as we only encode characters, that create compression in the first place. Therefore, we only store *runs* for single characters, that produce compression in the encoding, leaving the others untouched.

We also have the benefit of using nearly the whole range of possible *runs*(from 1 to 255). In the worst-case-scenario, we will just add the Comp_Bit_List(like in the 6th example), and nothing more, (unlike the other RLE-methods, who could and probably do, create even bigger data in the end), making the maximum overhead produced by Mespotine-RLE predictable. No matter how good the encoding produces compression or not: It's overhead, compared to the original-data, is never bigger than 32 Bytes!